\documentclass[]{aastex7}
\usepackage{amsmath}
\usepackage{gensymb}

\begin{document}

\title{Single Epoch Measurements of Dispersion Measure Gradients Towards PSR B0834+06}

\author[0000-0001-7888-3470]{Daniel Baker}
\affiliation{Academia Sinica Institute of Astronomy and Astrophysics
No.1, Sec. 4, Roosevelt Rd
Taipei 106319, Taiwan, R.O.C}
\email{dbaker@asiaa.sinica.edu.tw}
\author[0000-0003-2155-9578]{Ue-Li Pen}
\affiliation{Academia Sinica Institute of Astronomy and Astrophysics
No.1, Sec. 4, Roosevelt Rd
Taipei 106319, Taiwan, R.O.C}
\email{pen@asiaa.sinica.edu.tw}

\begin{abstract}
Understanding the evolution of pulsar dispersion measures is vital to high precision timing experiments, as well as astrometric experiments to determine pulsar positions and proper motions. In this work, we present a novel approach to measuring dispersion measure gradients using pulsar scintillometry. This approach makes use of the multipath propagation through the interstellar medium to simultaneously probe dispersion measures along multiple sight lines from a single observation. Using existing data of PSR B0834+06, we are able to measure gradients of $9.7\pm0.3\times10^{-6}~\rm{pc}~\rm{cm}^{-3}~\rm{mas}^{-1}$.
\end{abstract}


\section{Introduction} \label{sec:intro}
Pulsar emission is an important probe in understanding the free electron distribution of the Milky Way. The short duration and wide bandwidth of pulses makes them ideal for determining the Dispersion Measure (DM) of the Insterstellar Medium (ISM) along the line of sight. Due to the frequency-dependent index of refraction of the ionized plasma, the pulses arrive earlier at high frequencies and later at low frequencies. Since the DM depends only on the integrated electron column along the line of sight, these measurements can be combined with known pulsar distances to map the galactic electron distribution. Such measurements form the backbone of the electron density models from TC93 \citep{Taylor1993}, NE2001 \citep{Cordes2003}, and YMC16 \citep{Yao2017}.
As the pulsar moves across the sky, our line of sight intersects with different regions of the ISM resulting in changes in the electron column and DM. These variations introduce noise into precision timing experiments, such as Pulsar Timing Arrays (PTAs). Typically, this noise is reduced by observing PTA pulsars at higher frequencies, typically around $1.4~\rm{GHz}$, where the $\nu^{-2}$ frequency dependence of the delay reduces variability. However, this reduces the number of available pulsars as their steep spectra rapidly decrease S/N at higher frequencies. Understanding these variations could allow for the use of more pulsars at lower frequencies within PTAs. 

These DM monitoring experiments for many pulsars have been conducted from 150 MHz with LOFAR \citep{Donner2020} and the Parkes radio telescope \citep{Petroff2013} and many others. These studies show long term DM gradients as well as regular variation from the solar wind and month to years long fluctuations from smaller scale variations in the ISM. However, these experiemnts run up against two fundamental limitiation in their ability to model the ISM:
\begin{itemize}
    \item The cadence of the observations sets the minimum time scale of fluctuations that can be detected.
    \item They can only probe the direct line of sight to the pulsar as it travels across the sky, and so only give 1 one dimensional map
\end{itemize}

In addition to dispersive effects, the pulsar signal also experiences refraction in the ISM. Small scale changes in electron density result in refraction of the signal, and can result in it following multiple paths to the Earth. As a result, the observed pulses are convolved with the impulse response of the ISM. The interference of these different images causes variations in the pulsed intensity as a function of both time and frequency and is seen in the dynamic spectrum showing the time evolution of the amplitude of the frequency response (Fourier transform of the impulse response). This scattering also represents a nuisance for PTAs as it leads to a time variable pulse shape. However, as the different paths pass through different parts of the ISM, this scattering may also provide a new avenue to exploring DM variations on small scales.

The behavior of these different paths is more easily explored in the two dimensional Fourier Transform of the dynamic spectrum: the conjugate spectrum. In this space the interference between pairs of images are represented as single points corresponding to the differential time delay, $\tau$, and Doppler shift, $f_D$, of the two images. For many pulsars, the conjugate spectrum is dominated by scintillation arcs representing the interference of the line of sight with all scattered images, and inverted arclets showing the interference of pairs of scattered images. In order to study the behavior of individual images, it is helpful to solve the phase retrieval problem to recover the full complex frequency response as a function of time, known as the wavefield. The Fourier transform of the wavefield, called the conjugate wavefield for consistency with the conjugate spectrum, shows the delay and Doppler shift of each image relative to the line of sight as well as its complex magnification. Several approaches to solving this phase retrieval problem have been explored from variations on the CLEAN algorithm \citep{Walker2005}, to H-FISTA \citep{Oslowski2022}, and the $\theta-\theta$ transform \citep{Baker2021}. In many cases these scattered images appear to lie along a parabola in the $\left(\tau,f_D\right)$ space with
\begin{equation}
    \tau = \eta f_D^2
\end{equation}
where $\eta$ is reffered to as the arc curvature and encodes information about the orientation and velocities of the Earth,pulsar, and ISM. This parabolic strucutre implies a highly anisotropic scattering region with images forming along a nearly straight line on the sky.

These scattered images allow us to use the interstellar medium as a giant interferometer to perform precise astrometry on the pulsar. This has been used successfully to measure the size of the emission region of B0834 + 06 by \cite{Pen2014} and to determine the inclinations of pulsars in binary orbits (see \cite{Lyne1984}, \cite{Rickett2014}, and \cite{Reardon2020} for examples of several methods). The different images also open new pathways to study the behavior of the ISM at small ($\approx 1~\rm{AU}$) scales. \cite{Brisken2009} demonstrate that the scattering towards B0834+06 is dominated by a single thin, and highly isotropic, screen along the line of sight. This screen also interacts with a second isolated lensing location offset from the line of sight by nearly $25~\rm{mas}$. By examining the behavior of images scattered by both lenses, \cite{Zhu2023} show that the second lens is on the order $1~\rm{AU}$ in size and has properties consistent with being able to explain Extreme Scattering Events towards quasars. This opens the exciting possibility of using the behaviour of the images to probe changes in the ISM along nearby sightlines from a single observation. The most promising properties to study in this way are changes in the dispersion and rotation measures. In this work, we focus on the DM changes for features along the main arc. In principle these measurements can be combined with standard DM monitoring in order to construct a two dimensional map of the ISM, as the orientation of the screen need not be aligned with the motion of the pulsar.

In this paper, we explore two methods of detecting DM gradients using the frequency evolution of the images formed by a scintillation screen. In Sec.~\ref{sec:brisken} we discuss the detection of a DM gradient in the B0834+06 data of \cite{Brisken2009} by tracking individual features (Sec.~\ref{sec:brisken-image}) and by searching for delay shifts shifts in the arc entire arc (Sec.~\ref{sec:brisken-corr}). We then compare these to values to the gradients from some additional observations taken around the same time in Sec.~\ref{sec:archival}. Finally, the connection to other observables is explored in Sec.~\ref{sec:other_observables} and the general ramifications in Sec.~\ref{sec:ramifications}.

\section{Brisken Data}
\label{sec:brisken}
\begin{figure}[ht!]
    \centering
    \includegraphics{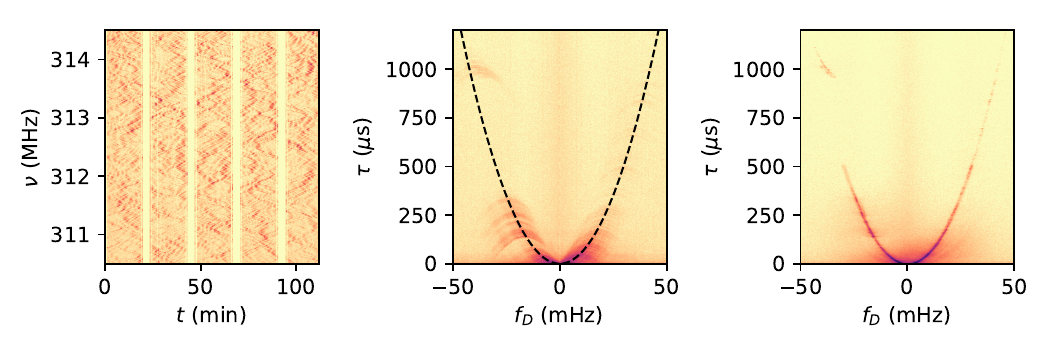}
    \caption{Portion of the Arecibo dynamic spectrum (left), its conjugate spectrum, and recovered wavefield from \cite{Brisken2009}. }
    \label{fig:brisken-spectra}
\end{figure}
We begin by reanalyzing the data taken by \cite{Brisken2009} of B0834+06 that were used to create the first VLBI image of a one-dimensional scattering screen. For this work, we focus on the single station Arecibo data taken on 2005 November 12. Dynamic spectra were originally created using the DiFX correlator with four $8~\rm{MHz}$ sub-bands from $310.5$ to $342.5~\rm{MHz}$ with a total of 131072 channels using $1.25~\rm{s}$ integrations.  For this work, we rebinned the dynamic spectra by a factor of 4 in time to improve performace of phase recovery with the $\theta-\theta$ transform. Wavefield recovery was achieved using the $\theta-\theta$ transform method as described in \cite{Baker2021}. The dynamic spectrum is broken into a series of small overlapping chunks in time and frequency and a $\theta-\theta$ matrix is generated for each one. Using an eigenvector decomposition, the complex magnifications along the bottom of the main arc are determined and the wavefield recovered within each chunk. These wavefields are then given a complex rotation and stacked such that the phases agree within the overlapping areas. Finally, the square-root of the amplitudes from the original dynamic spectrum are used to set the wavefield amplitudes to recover more images in the conjugate wavefield. An example of a portion of the dynamic spectrum as well as its conjugate spectrum and corresponsing wavefield are shown in Fig.~\ref{fig:brisken-spectra}. The off arc feature at $1~\rm{ms}$ has been discussed in detail in the original \cite{Brisken2009} paper as well as in \cite{Zhu2023}. In this work we will focus on features along the main arc; whose geometry, as measured in \cite{Liu2016} is given in. Most importantly for our purposes is the screen distance of $389\pm5~\rm{pc}$ and orientation of $154\pm5\degree$

\subsection{Image Tracking}
\label{sec:brisken-image}
\begin{figure}[htb!]
    \centering
    \includegraphics{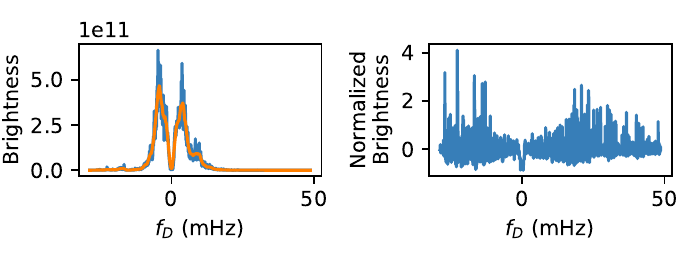}
    \caption{Average brightness along the main arc as a function of $f_D$. The left panel shows the full average and the $1.3~\rm{mHz}$ smoothed version, while the right panel shows the normalized brightness as described in the text.}
    \label{fig:brisken-brightness}
\end{figure}

\begin{figure}[ht!]
    \centering
    \includegraphics{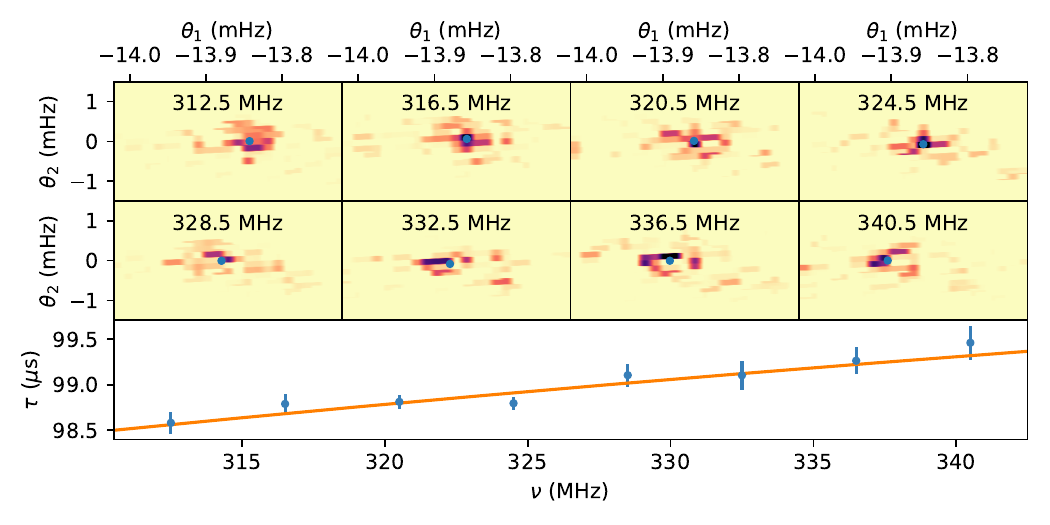}
    \caption{Frequency evolution of a single image in $\theta-\theta$ space (upper panels) with the fitted location at the frequency marked in blue and the corresponding $\tau$ coordinates (bottom) with the best fit DM delay curve in orange.}
    \label{fig:brisken-single-image}
\end{figure}

Since the recovered wavefield shows many isolated features, particularly towards the ends of the arc, we first check for signs of a DM gradient by tracking the position of those features over frequency. The wavefield is divided into 8 sub-bands of $4~\rm{MHz}$ as in Fig.~\ref{fig:brisken-spectra} and a $\theta-\theta$ matrix formed for each and correcting for the frequency dependence of $f_D$ by rescaling all $f_D$ values to $326.5~\rm{MHz}$. The $\theta-\theta$ matrices were then cut down to cover from $-30$ to $50~\rm{mHz}$ in $\theta_1$, to cover the majority of the main arc and avoid the millisecond feature, and from $-1.5$ to $1.5~\rm{mHz}$ in $\theta_2$ to only consider points close to the main arc. The amplitudes of the $\theta-\theta$ matrices were averaged over $\theta_2$ and frequency to produce an average brightness in which to search for images. To correct for the large-scale evolution of the background, a $1.3~\rm{mHz}$ wide top hat was convolved with the brightness to produce a smooth background as shown in Fig.~\ref{fig:brisken-brightness}. A ``normalized brightness" is produced by dividing by this smoothed brightness and subtracting $1$.
To extract images from the normalized brightness, we first locate its maximum and the first zero crossing to the left and right and fit a Gaussian peak to this region. This peak is then subtracted from the normalized brightness, and we repeat the process until no values greater than $1$ remain. For this observation, we detect approximately $60$ images using this method. 
To track each of these images, we first select a region of the $\theta-\theta$ matrix around the image in each band and average over all frequencies. A two-dimensional Gaussian is then fitted to determine a reference location. We then convolve this average with the image region for each sub-band and perform a second fit to determine the offset from the mean. The resulting positions for an example image is shown in Fig.~\ref{fig:brisken-single-image} as well as the corresponding $\tau$ coordinates. The delay evolution is then fit to
\begin{equation}
    \tau = \tau_\infty + \frac{\Delta\rm{DM}}{K}\frac{1}{\nu^2}
\end{equation}
Where the delay at infinite frequency, $\tau_\infty$, and $\Delta\rm{DM}$ are to be fit for; and $K$ is the dispersion constant
\begin{eqnarray}
    K &=& \frac{1}{8\pi^2}\frac{e^2}{\epsilon_0m_ec}\times \rm{parsec}\\
    &=&241.0331786~\rm{pc}~\rm{cm}^{-3}~\rm{GHz}^{-2}\rm{s}^{-1}
\end{eqnarray}
In order to be sure we are tracking actual images, we first note that $\tau_{\infty}$ represents the position of the image with no dispersive effects and so should fall along the parabolic arc at our reference frequency. We therefore discard all points where $\left|\tau_\infty - \eta_{\rm{ref}}f_D^2\right|>100~\mu\rm{s}$ and any with large $\chi^2_{\rm{red}}$. This leaves us with 39 out of our original 57 images.

\begin{figure}[ht!]
    \centering
    \includegraphics{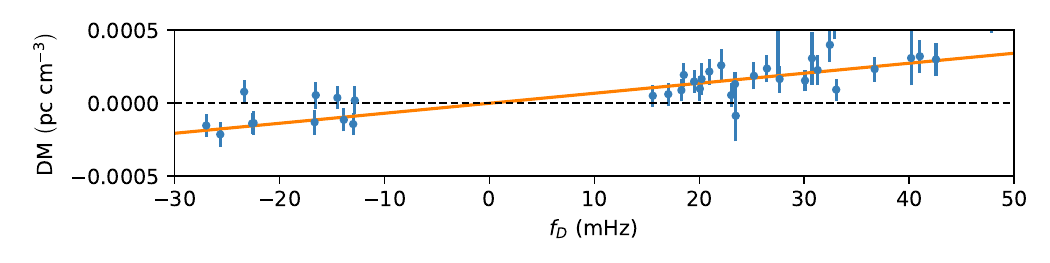}
    \caption{Dispersion Measure for each of the tracked images as a function as $f_D$ (blue) and the best fit gradient $6.9\pm0.7\times10^{-6}~\rm{pc}~\rm{cm}^{-3}~\rm{mHz}^{-1}$.}
    \label{fig:brisken_feature_track_grad}
\end{figure}

Plotting the recovered $\Delta\rm{DM}$ against position along the arc in Fig.~\ref{fig:brisken_feature_track_grad} shows a clear trend. To capture the general trend, we fit a simple DM gradient along the arc. However, the spread of DMs around this gradient is larger than expected from the errors of the individual fits. This may be caused by small-scale variations in DM on top of the global gradient. In order to incorporate this variation into our fit, we add a constant $6\times10^{-5}~\rm{pc}~\rm{cm}^{-3}$ to all errors to achieve $\chi^2_{\rm{red}}=1$. This yields a DM gradient across the arc of $6.9\pm0.7\times10^{-6}~\rm{pc}~\rm{cm}^{-3}~\rm{mHz}^{-1}$.

\subsection{Correlation Analysis}
\label{sec:brisken-corr}
Since not all wavefields have such clear isolated images for us to track, other methods that do not rely on individual features will in general be required. One possibility is to look for the evolution of individual $f_D$ bins over frequency (after correcting for the curvature evolution of the arc). 

\begin{figure}
    \centering
    \includegraphics{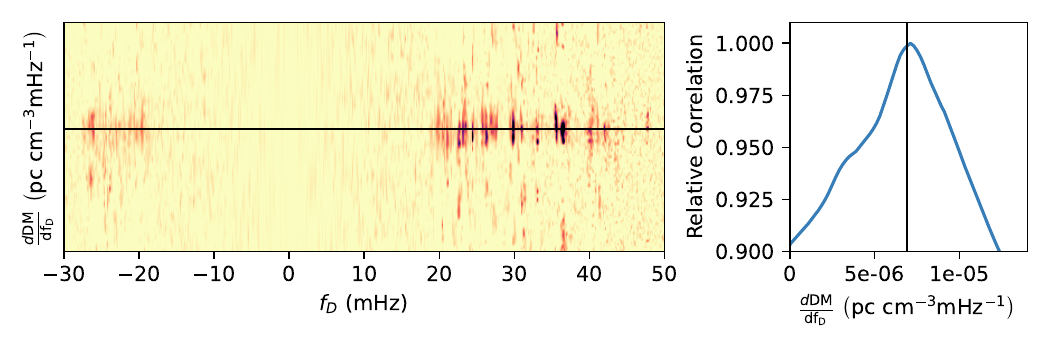}
    \caption{Correlation between the arc at $320.5~\rm{MHz}$ and $312.5~\rm{MHz}$ for each $f_D$ (left) and averaged over all $f_D$ (right). The black lines indicate the best fit value of the DM gradient from the feature tracking method.}
    \label{fig:GB057_correlation}
\end{figure}

To do this, we first mask out the region below $50~\mu\rm{s}$ in the wavefield where the reconstruction is blurred. We then create a grid of DM gradients between $-1\times10^{-4}~\rm{pc}~\rm{cm}^{-3}~\rm{mHz}$ and $1\times10^{-4}~\rm{pc}~\rm{cm}^{-3}~\rm{mHz}$ and calculate the corresponding time delay between each frequency band and the first band for each $f_D$ bin. Each bin is then shifted by the corresponding delay and correlated against the lowest frequency band. Fig.~\ref{fig:GB057_correlation} shows an example of this correlation for each $f_D$ along the main arc, as well as the average over all $f_D$ for a single band. 
We then fit a Gaussian to the peak of the average correlation at each frequency. We estimate the error on each fit by taking the standard deviation of the residuals and determining the region for which the correlation is within one standard deviation of the fitted peak. Fitting a constant DM gradient over frequency using these results gives a final measurement of $7.0\pm0.2\times 10^{-6}~\rm{pc}~\rm{cm}^{-3}~\rm{mHz}^{-1}$. This agrees nicely with the results from the individual image tracking approach, which provides a nice cross check for the two methods.

\section{Archival Arecibo Data}
\label{sec:archival}
\begin{figure}[ht!]
    \includegraphics{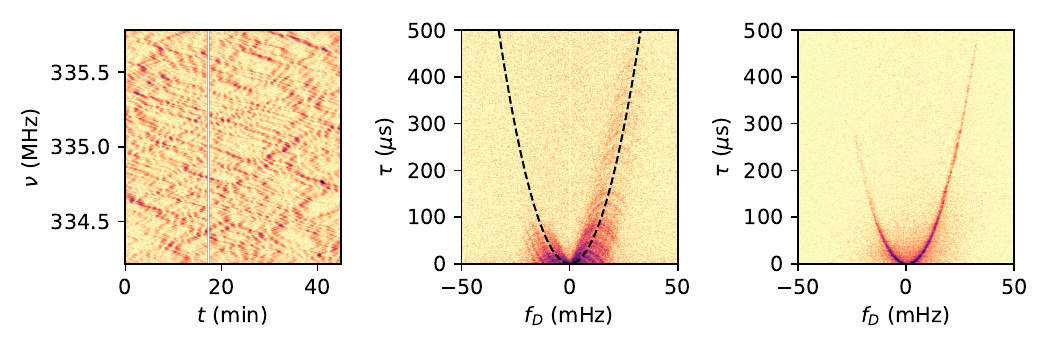}
    \caption{Example dynamic spectrum (left), conjugate spectrum (center), and recovered conjugate wavefield for the observation taken at 335MHz on MJD 53648. The best fit curvature is marked in black on the conjugate spectrum.}
    \centering
    \label{fig:treasure_spec}
\end{figure}
In addition to the Brisken data, we also make use of eight additional observations of B0834+06 taken at Arecibo from October to December 2005. Observations were taken simultaneously at three bands centered at $317$, $318.5$, and $335~\rm{MHz}$. However, the frequency resolution of the $318.5~\rm{MHz}$ data is different than the other two bands so we focus on them. This also has the advantage of focusing on the larger frequency difference making changes in DM more obvious. The dynamic spectra for these bands are formed with $10~\rm{s}$ sub integrations and $763~\rm{Hz}$ channels covering a bandwidth of $1.5~\rm{MHz}$ for between $30$, and $40$ minutes. They were previously used in \cite{Zhu2023} in order to study a potential extreme scattering event. An example of the dynamic and conjugate spectra as well as the conjugate wavefield for the observation on MJD 53648 is shown in Fig.~\ref{fig:treasure_spec}. These data are expected to be part of a larger data release of archival Arecibo dynamic spectra later this year.

When running the image tracking approach on these observations, we were unable to consistently identify enough points to track for a gradient fit and so we instead focus on the correlation approach. As with the Brisken data, we scale all $f_D$ values to $326.5~\rm{MHz}$. The resulting gradients seem broadly consistent with the results from the Brisken data. However, since the $f_D$ corrdinate of the images depends on both the angular offset of the image and the effective velocity, which varies over the Earth's orbit, a full comparison requires that we convert these into angular corrdinates on the sky for each day. For a given image, the angular offset $\theta$ along the screen and the Doppler shift are related by
\begin{equation}
    f_D = \frac{v_\parallel}{\lambda}\theta
\end{equation}
for a given observing wavelength $\lambda$. Here, the effective parallel velocity is given by 
\begin{equation}
    v_\parallel = \left(\boldsymbol{v}_{\oplus} + \boldsymbol{v}_p \frac{d_s}{d_p-d_s} - \boldsymbol{v}_{\text{ISM}}\frac{d_p}{d_p-d_s}\right)\cdot \boldsymbol{n}
\end{equation}
where $\boldsymbol{v}_{\oplus}$, $\boldsymbol{v}_p$, and $\boldsymbol{v}_{\text{ISM}}$ are the velocities of the Earth, pulsar, and ISM respectively; while $d_s$, and $d_p$ are the distances to the screen and pulsar and $\boldsymbol{n}$ is the direction along the screen. For the main Brisken observation this yields a spatial gradient of $9.7\pm0.3\times10^{-6}~\rm{pc}~\rm{cm}^{-3}~\rm{mas}^{-1}$.
\begin{figure}[!ht]
    \centering
    \includegraphics{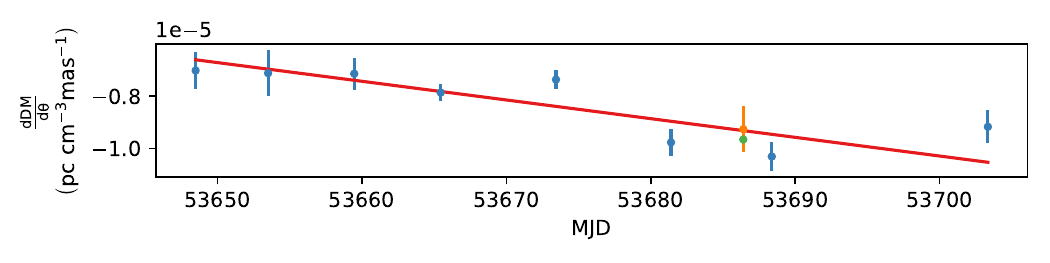}
    \caption{Fitted DM gradients along the screen for several days of observations. Blue points show the additional observations discussed in this section, while the green and orange points show the results from the Brisken data using the image tracking and correlation methods respectively. The red line shows the best fit linear trend to the correlation results for both datasets with a slope of $-2.6\pm0.4\times10^{-5}~\rm{pc}~\rm{cm}^{-3}~\rm{mas}^{-1}~\rm{yr}^{-1}$}
    \label{fig:dm_grad_evo}
\end{figure}

There appears to be a slight linear trend to the gradients over these observations which we fit to have a slope of $-2.6\pm0.4\times10^{-5}~\rm{pc}~\rm{cm}^{-3}~\rm{mas}^{-1}~\rm{yr}^{-1}$.

\section{Relation to other observables}
\label{sec:other_observables}
Spatial DM gradients can manifest as changes in several other potential observables in Pulsars. We consider several such effects here. In general, these estimates will represent lower bounds, as our measurements only give the DM gradients along the main axis of the scintillation screen.

The most obvious of the effects is that DM gradients act like prisms causing a chromatic variation in the apparent position of the pulsar. Assuming the refraction is dominated by a thin region along the line of sight the angular offset $\Delta\theta$ is given by
\begin{equation}
    \label{eq:delta_theta}
    \Delta\theta = \frac{1}{d_{\rm{eff}}}\frac{\lambda^2}{2\pi}r_e\frac{dDM}{d\theta}
\end{equation}
where the effective distance $d_{\rm{eff}}$ is given by $\frac{d_{p}d_{s}}{d_{p}-d_{s}}$, and $r_e$ is the classical electron radius. This  results in an offset of $-0.149\pm0.004~\rm{mas}$ for the Brisken observation. Additionally, if the observed change in DM gradients persists then the apparent proper motion of the pulsar from parallax measurements would be off by $-0.40\pm0.06~\rm{mas}~\rm{yr}^{-1}$ along the screen direction. 

Another consideration is that the wavelength dependence in Eq.~\ref{eq:delta_theta} meand that the pulsar position position will shift with frequency. This results in two additional frequency dependent contributions to the delay: a geometric delay due to the additional path length, and a variable dispersion measure as the different lines of sight pass through different points along the DM gradient. The geometric delays is given by
\begin{eqnarray}
    \tau_{\rm{geo}} &=& \frac{\Delta\theta^2d_{\rm{eff}}}{2c}\\
    &=& \frac{\left(r_e\frac{dDM}{d\theta}\right)^2}{8\pi^2d_{\rm{eff}}\cdot c}\lambda^4,
\end{eqnarray}
while the additional dispersive delay is
\begin{eqnarray}
    \tau_{\rm{disp}} &=& \frac{1}{K}\Delta\theta\frac{dDM}{d\theta}\left(\frac{\lambda}{c}\right)^2\\
    &=& \frac{1}{K}\frac{r_e\left(\frac{dDM}{d\theta}\right)^2}{2\pi d_{\rm{eff}} c^2}\lambda^4
\end{eqnarray}
Interestingly, both terms are quadratic in the the DM gradient and quartic in wavelength or frequency. This would appear as a frequency dependent DM: $DM = DM_{\infty} + \Delta DM\left(\nu\right)$ where
\begin{equation}
    \Delta DM\left(\nu\right)=\frac{c^3r_e}{2\pi d_{\rm{eff}}}\left(\frac{dDM}{d\theta}\right)^2\left(\frac{r_e~K}{4\pi}+\frac{1}{c}\right)\nu^{-2}
\end{equation}
For the relatively narrow band of the Brisken observation, this could result in an offset in the arrival time between the bottom and top of the band of between 20 to 40 ns after dedispersion depending on which DM was used. For wide band observations, the difference, for this DM gradient, could be nearly 200 ns between 300 and 1600 MHz. Since both $d_{\rm{eff}}$ and $\frac{dDM}{d\theta}$ are expected to scale with pulsar distance,and therefore approximately with DM, We expect this effect to become more pronounced for higher DM pulsars. 

Finally, since DM gradients are typically measured in the time domain we convert into a time derivative by noting that features move through the wavefield at a predictable rate of
\begin{equation}
    \dot f_D = \frac{1}{2\eta\nu}
\end{equation}
where $\eta$ is the arc curvature. This gives us a gradient of $5.0\pm0.1\times10^{-4}~\rm{pc}~\rm{cm}^{-3}~\rm{yr}^{-1}$. As mentioned before, this gradient represents the contribution due to the projected motion along the line of scattered images. When combined with more traditional approaches using regular monitoring of the DM over several epochs, which only determines the gradient along the pulsar's proper motion, one could determine the full two dimensional gradient on the sky.

\section{Ramifications}
\label{sec:ramifications}
This paper presents two related techniques for measuring spatial DM gradients using pulsar scintillometry using either the behavior of individual scattered images, or using the average behavior of many images. We demonstrate how for observations of PSR B0834+06 where individual images are traceable, both methods are consistent and alow us to measure angular gradients of $9.7\pm0.3~\rm{pc}~\rm{cm}^{-3}~\rm{mas}^{-1}$. We also examine several additional observations taken around the same date and see boradly consistent results that evolve slowly over several weeks.

These new techniques offer several advantages over traditional approaches to studying DM trends. Firstly, we can measure the gradients from a single observation by making use of the multipath propagation through the ISM to probe multiple sight lines simultaneously. Potentially reducing the observational cadence required to track DM since we only need one observation per size of the scattering region. In the case where individual images are trackable, we can detect their individual DMs allowing us to investigate small scale variations on top of the overall gradient.

Since these spatial gradients can induce a frequency dependency on the DM, measuring them allows us the remove this effect and improve the measurement of the direct line of sight DM which we call $DM_{\infty}$. Failing to correct for these gradients can result in timing offsets of up to $200~\rm{ns}$ at $1600~\rm{MHz}$ for B0834+06 depending on the frequency for which DM is measured. For pulsars with larger DMs, we can expect this delay to scale accordingly. 

These spatial gradients will also result in refractive shifts in the apparent pulsar position with frequency. Variations in this offset introduce an extra source of noise into VLBI measurements of the pulsar position and velocity over time and, in cases such as the one explored here where there is an approximately constant second derivative in the electron column, yield a persistent bias. While measurements at higher frequencies can reduce this effect, as in the PSR$\pi$ project \citep{Deller2016}, measurement of these small scale gradients and correcting for them will open the possiblity to study more pulsars in the lower frequency regime.

Going forward, these measurements of the gradient along the scattering direction can be combined with traditional DM monitoring to determine the two dimensional gradient of the DM on the sky to help correct for refractive effects in timing and astrometric observations as well as study the small scale structure of the ISM.

\begin{acknowledgments}The authors would like to thank Dan Stinebring and Walter Brisken for the use of their data.\end{acknowledgments}

\bibliography{sample631}{}
\bibliographystyle{aasjournal}

\end{document}